\documentstyle[preprint,epsfig,aps]{revtex}

\begin{document}
\draft
\title{Evidence for Exotic Meson Production  in the Reaction
$ \pi^{-} p \rightarrow \eta \pi^{-} p$ 
at 18  $ {\rm GeV}/c$}

\author{D.~R.~Thompson\rlap,$^1$
G.~S.~Adams\rlap,$^7$
T.~Adams\rlap,$^1$ 
Z.~Bar-Yam\rlap,$^4$
J.~M.~Bishop\rlap,$^1$ 
V.~A.~Bodyagin\rlap,$^5$
D.~S.~Brown\rlap,$^6$
N.~M.~Cason\rlap,$^1$ 
S.~U.~Chung\rlap,$^2$ 
J.~P.~Cummings\rlap,$^{4}$
S.~P.~Denisov\rlap,$^3$
V.~A.~Dorofeev\rlap,$^3$
J.~P.~Dowd\rlap,$^4$
P.~Eugenio\rlap,$^4$
R.~W.~Hackenburg\rlap,$^2$ 
M.~Hayek\rlap,$^4$
E.~I.~Ivanov\rlap,$^1$
I.~A.~Kachaev\rlap,$^3$
W.~Kern\rlap,$^4$
E.~King\rlap,$^4$
O.~L.~Kodolova\rlap,$^5$
V.~L.~Korotkikh\rlap,$^5$
M.~A.~Kostin\rlap,$^5$
J.~Kuhn\rlap,$^7$
V.~V.~Lipaev\rlap,$^3$
J~.M.~LoSecco\rlap,$^1$
J.~J.~Manak\rlap,$^1$
J.~Napolitano\rlap,$^7$
M.~Nozar\rlap,$^7$
C.~Olchanski\rlap,$^2$
A.~I.~Ostrovidov\rlap,$^5$
T.~K.~Pedlar\rlap,$^6$
A.~V.~Popov\rlap,$^3$
D.~I.~Ryabchikov\rlap,$^3$
 A~.H.~Sanjari\rlap,$^1$ 
L.~I.~Sarycheva\rlap,$^5$
K.~K.~Seth\rlap,$^6$
W.~D.~Shephard\rlap,$^1$ 
N.~B.~Sinev\rlap,$^5$
J.~A.~Smith\rlap,$^7$
D.~L.~Stienike\rlap,$^1$ 
C.~Strassburger\rlap,$^{2,}$\cite{byline2}
S.~A.~Taegar\rlap,$^1$
I.~N.~Vardanyan\rlap,$^5$
D.~P.~Weygand\rlap,$^2$
D.~B.~White\rlap,$^7$
H.~J.~Willutzki\rlap,$^2$
J.~Wise\rlap,$^6$
M.~Witkowski\rlap,$^7$
A.~A.~Yershov\rlap,$^5$
D.~Zhao\rlap,$^6$
}
\author{The E852 Collaboration}
\address{$^1$University of Notre Dame, Notre Dame, IN 46556, USA\\
$^2$Brookhaven National Laboratory, Upton, Long Island, NY 11973, USA\\
$^3$Institute for High Energy Physics, Protvino, Russian Federation\\
$^4$University of Massachusetts Dartmouth, North Dartmouth, MA 02747, USA\\
$^5$Moscow State University, Moscow, Russian Federation\\
$^6$Northwestern University, Evanston, IL 60208, USA\\
$^7$Rensselaer Polytechnic Institute, Troy, NY 12180, USA}
\date{\today}

\maketitle
\vskip12cm
\begin{abstract}
The $\eta \pi^{-}$ system has been 
studied in the reaction
$\pi^{-} p \rightarrow \eta \pi^{-} p$ at $18$ 
${\rm GeV}/c$.  A large asymmetry 
 in the angular distribution is observed  indicating
interference between L-even and L-odd partial waves. The $a_{2}(1320)$ 
is observed 
in the $J^{PC}$ = $2^{++}$ wave, 
as is a broad 
enhancement between 1.2 and 1.6 ${\rm GeV}/c^{2}$ in the  
$1^{-+}$ wave.  The  
observed 
phase difference between 
these  waves shows that there is 
phase motion in addition to that due to
 $a_{2}(1320)$ decay.
The data can be fitted by interference between
the $a_{2}(1320)$ and an exotic $1^{-+}$ resonance with
M = (1370  $\pm16$ ${+50}\atop{-30}$) ${\rm MeV}/c^2$ and 
$\Gamma =$
(385  $\pm40$ ${+65}\atop{-105}$) ${\rm MeV}/c^2$.
\end{abstract}

 
\narrowtext


The question of whether or not 
hadrons outside the scope of the constituent 
quark model exist is one
whose answer speaks directly to the fullness of our 
 understanding of quantum chromodynamics 
(QCD) \cite{th:ikp}.  However, non-$q\overline{q}$ mesons
(or exotic mesons) 
have proven difficult to distinguish from the 
many conventional $q\overline{q}$ states which 
populate the various mesonic spectra.  For this
reason, much attention has been focused on those 
states with manifestly exotic $J^{PC}$ quantum numbers.

A $q\overline{q}$ meson with orbital angular momentum
$\ell$ and total spin $s$ must have $P = (-1)^{\ell+1}$
and $C = (-1)^{\ell+s}$.  Thus a resonance with 
$J^{PC}$ = $0^{--}$, $0^{+-}$, $1^{-+}$, $2^{+-}$, ...
must be exotic. 
Such a state could be a gluonic  excitation such
as a hybrid ($q\overline{q}g$) or glueball ($2g, 3g, ...$), 
or a multiquark ($q\overline{q}q\overline{q}$) state.
In a relative P wave (L=1), the $\eta \pi^{-}$ 
system has $J^{PC}$ = $1^{-+}$. 
Having isospin I=1, it could not be a glueball, but it could
be  a hybrid or a multiquark state.

Production and decay properties of exotic 
states have been predicted using several models 
\cite{th:ip,th:barnes,th:clp,th:bcs,th:chanowitz,jaffe,lattice1}. 
A calculation based upon the MIT bag model predicts
\cite{th:barnes} that a  $1^{-+}$ hybrid ($q\overline{q}g$)
will have a mass near 1.4 ${\rm GeV}/c^{2}$.
On the other hand, the flux-tube model \cite{th:clp,th:bcs}
predicts the mass 
of the lowest-lying hybrid state to be
around $1.8$ ${\rm GeV}/c^{2}$. 
Characteristics of bag-model S-wave multiquark
states (which would have $J^{P}$ = $0^{+}$, 
$1^{+}$, or $2^{+}$) have been predicted\cite{jaffe} but
those for a $1^{-}$ state have not. 
Finally, 
recent lattice calculations\cite{lattice1} of the $1^{-+}$
hybrid meson estimate its mass to be in the range
of 1.7 to 2.1 GeV. 

The $\eta \pi$ system has been studied in several recent 
experiments, with apparently inconsistent results.  Alde et al. 
\cite{ex:GAMSetapz}, in a study of $\pi^{-}p$
interactions at 100 GeV/$c$ at CERN
(the GAMS experiment),
claimed to observe a $1^{-+}$ 
state  in the 
$\eta \pi^{0}$ system at 1.4 ${\rm GeV}/c^{2}$ 
produced via \em unnatural \rm parity exchange 
(the P$_{0}$ partial wave---the naming convention is discussed below) 
\cite{prokosh}.  Aoyagi et al. \cite{ex:KEKetapm}, 
in a $\pi^{-}p$ experiment at 6.3 GeV/$c$ at
 KEK,  observed a rather narrow enhancement
in the $\eta \pi^{-}$ system
at 1.3 ${\rm GeV}/c^{2}$
in the \em natural\rm~parity exchange $1^{-+}$ spectrum (P$_{+}$).  
Beladidze et al. \cite{ex:VESetapm}, in the VES experiment
at IHEP,  
($\pi^{-}N$ interactions at 37 GeV/$c$)
also reported a  
P$_{+}$ signal in the $\eta \pi^{-}$ state, but their
signal  was broader and  had a significantly
different  phase variation from that of the KEK experiment.
While the phase difference between the P$_{+}$ and D$_{+}$ waves
was independent of $\eta \pi$ mass in the KEK analysis, 
that phase difference did show significant mass dependence
in the VES analysis.  (Since the phase variation for the
D$_{+}$ wave follows a classic Breit-Wigner
pattern for the $a_{2}(1320)$ meson, the phase difference
between these waves can determine the phase variation of
the unknown P$_{+}$ wave.)

Here we study the $\eta \pi^{-}$ system in the reaction 
$  \pi^{-} p \rightarrow \eta \pi^{-} p$ 
at $18$ ${\rm GeV}/c$.  Our data sample was 
collected in the first data run of E852 
at the AGS at
Brookhaven National Laboratory
with the Multi-Particle Spectrometer (MPS) 
\cite{mps:ozaki40}
using a liquid hydrogen target.  The MPS, which
was equipped with six drift-chamber
modules  \cite{nim:drift} and three
proportional wire chambers,
was augmented by: a
four-layer cylindrical drift chamber surrounding the target 
\cite{nim:tcyl};  a soft-photon detector consisting
of 198 blocks of thallium-doped cesium iodide 
\cite{nim:csi} also surrounding the target;
a window-frame lead-scintillator photon-veto counter;
a large drift chamber; and a 3045-element 
lead-glass detector (LGD)  \cite{nim:lgd}
downstream of the MPS.  Further  details 
 are given elsewhere 
 \cite{teige}.

A total of 47 million  triggers 
which required one  forward-going charged track, one recoil
charged track, and an LGD trigger-processor signal
enhancing high electromagnetic effective mass was recorded.
Of these, 47,200 events were reconstructed which
were consistent with 
the $\eta\pi^-p ~(\eta\rightarrow2\gamma)$ final state.  
These events satisfied topological and fiducial volume
cuts, as well as energy/momentum conservation 
for production and for the  $\eta\rightarrow2\gamma$
decay with a confidence level $>10$\%\cite{squaw}.  The
 $2\gamma$ mass resolution at the $\eta$
mass is $\sigma$ = 0.03 ${\rm GeV}/c^2$.

\begin{figure}
\begin{center}
\mbox{\epsfig{file=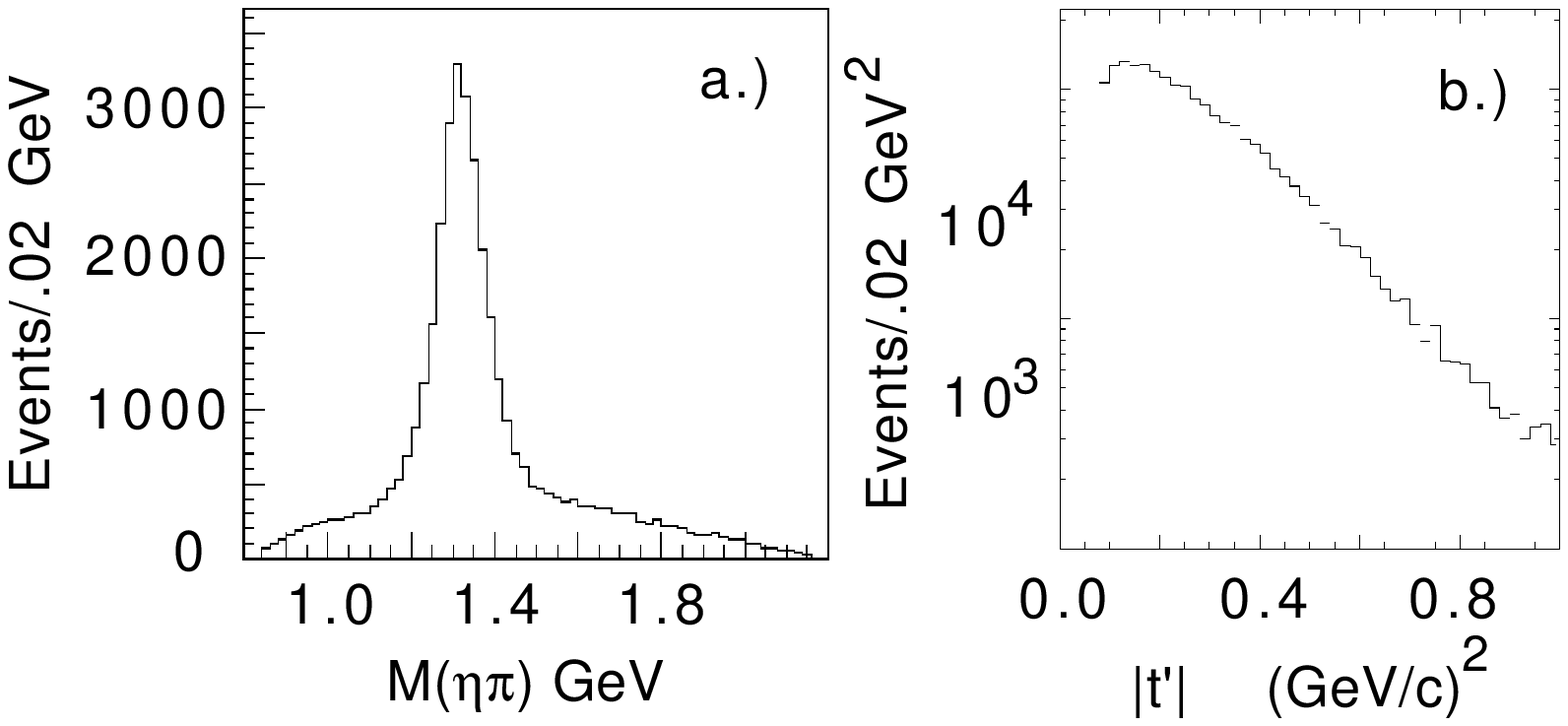,width=6.5in}}\vskip15mm
\end{center}
\caption{a.)  The $\eta\pi^-$ effective mass distribution. 
    b.) Distribution of $|t^{\prime}|= |t|-|t|_{\rm{min}}$.
}
 \label{f:mass}
 \end{figure}

The $a_{2}(1320)$ is the dominant feature of the 
$\eta \pi^{-}$ mass spectrum shown in Fig.\,\ref{f:mass}a.
Background has been estimated using
side bands in both the 2-$\gamma$ mass distribution and the
missing-mass distribution, thus  
taking into account background from non-$\eta$ sources as well as
from sources due to production of other final states.  The background
level is approximately 7\% at 1.2 ${\rm GeV}/c^2$, falling to 1\%
at 1.3 ${\rm GeV}/c^2$.

The acceptance-corrected distribution of 
$|t^{\prime}|= |t|-|t|_{\rm{min}}$,
where
t is the 
the four-momentum-transfer, is shown for 
$|t^{\prime}|> 0.08(\rm {GeV}/c)^2$ in 
Fig.\,\ref{f:mass}b.  (Our acceptance is quite low below 0.08
 $(\rm {GeV}/c)^2$ due to a trigger requirement.) 
The shape of this distribution is consistent with previous
experiments and has been shown to be  consistent with  
natural-parity exchange production in 
Regge-pole phenomenology \cite{th:regge,th:sacharidis}.

\begin{figure}
\hglue0cm
\hbox{\vrule height12.5pt depth3.5pt width0pt}
\noindent
\mbox{\epsfig{file=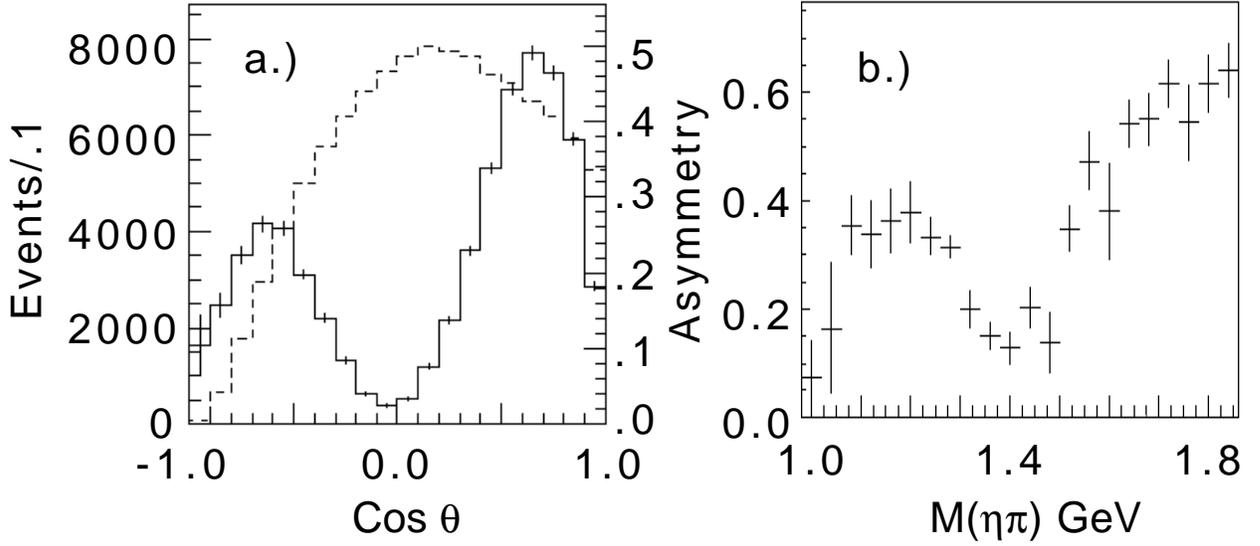,width=6.9in}}\vskip15mm
 \caption{Distributions of a.) the cosine of the decay angle
in the Gottfried-Jackson frame
for events with 
1.22 $<M(\eta \pi^{-})<$1.42 ${\rm GeV}/c^{2}$, and  b.) the 
forward-backward decay asymmetry as a function of $M(\eta \pi^{-})$.
The asymmetry $=(F-B)/(F+B)$ where F(B) is the
number of events for which the $\eta$'s momentum is forward 
(backward) 
in the Gottfried-Jackson frame.   The dashed curve
and the right-hand scale
in a.) show the acceptance 
in this mass region.}
 \label{f:cos}
 \end{figure}

The acceptance-corrected distribution of
$\cos\theta$, the cosine of the angle between 
the $\eta$ and the beam track 
in the Gottfried-Jackson frame \cite{GJframe} of the $\eta\pi^-$ 
system, is  shown in Fig.\,\ref{f:cos}a 
for $1.22 < M(\eta \pi^{-}) < 1.42$ ${\rm GeV}/c^{2}$.
There is  a  forward-backward asymmetry in 
$\cos\theta$.  
The asymmetry 
for $|\cos\theta| < 0.8$ is plotted as a function
of $\eta \pi^{-}$ mass 
in Fig.\,\ref{f:cos}b.
The asymmetry is large, statistically significant and
mass dependent.  
Since the presence of only even values of L would yield 
a symmetric distribution in $\cos\theta$,
the observed asymmetry requires that odd-L 
partial waves be present to describe
the data.

A partial-wave analysis (PWA) \cite{th:SUform,th:SUtwo} 
based on the extended maximum
likelihood method has been used to study the 
spin-parity structure of the $\eta \pi^{-}$ system.  
The partial waves are parameterized in terms of the
quantum numbers $J^{PC}$ as well as $m$, the {\em absolute 
value} of the angular momentum projection, and the reflectivity 
$\epsilon$ (which is positive (negative) for natural (unnatural)
parity exchange
\cite{th:SUTr}).  
In our naming convention, a letter indicates the angular
momentum of the 
partial wave in standard spectroscopic notation, while
a subscript of $0$ means $m$ = $0$, $\epsilon$ = $-1$, 
and a subscript of $+(-)$ means $m$ = $1$, $\epsilon$ = $+1(-1)$.
Thus, S$_{0}$ denotes the partial wave having
$J^{PC}m^{\epsilon}$ = $0^{++}0^{-}$, while P$_{-}$ 
signifies $1^{-+}1^{-}$, D$_{+}$ means $2^{++}1^{+}$,
and so on.  We consider  partial waves with $m \leq 1$, 
and we assume that the production
spin-density matrix has rank one.

The experimental acceptance is determined by a 
Monte Carlo method.  Peripherally-produced events are generated 
 \cite{sage} with isotropic angular distributions
in the Gottfried-Jackson frame.
After adding detector simulation \cite{geant}, 
the Monte Carlo
event sample is subjected to the same event-selection cuts 
and run through the same analysis as
the data.  The
experimental acceptance is then incorporated into the PWA by
using these events to calculate 
normalization integrals (see ref. \cite{th:SUform}).

Goodness-of-fit is determined by calculation 
of a $\chi^{2}$ from comparison 
of the experimental moments with those 
predicted by the results of the PWA fit.  
A systematic study has been performed to determine the 
effect on goodness-of-fit of adding and subtracting 
partial waves of $J \leq 2$ and $m \leq 1$.  
All such waves have been included in the final fit.
We have also performed fits including partial waves with $J$ = $3$
and $J$ = $4$.
Contributions from these partial waves are found to be insignificant 
for $M(\eta \pi^{-}) < 1.8$ ${\rm GeV}/c^{2}$.
Thus,  PWA fits shown or referred to in this 
letter include all partial waves with $J \leq 2$ and $m \leq 1$
(i.e. S$_{0}$, P$_{0}$, P$_{-}$, D$_{0}$, D$_{-}$, P$_{+}$, 
and D$_{+}$).  The background described above was included as a
non-interfering, isotropic
term of fixed magnitude. 

\begin{figure}
\begin{center}
\mbox{\epsfig{file=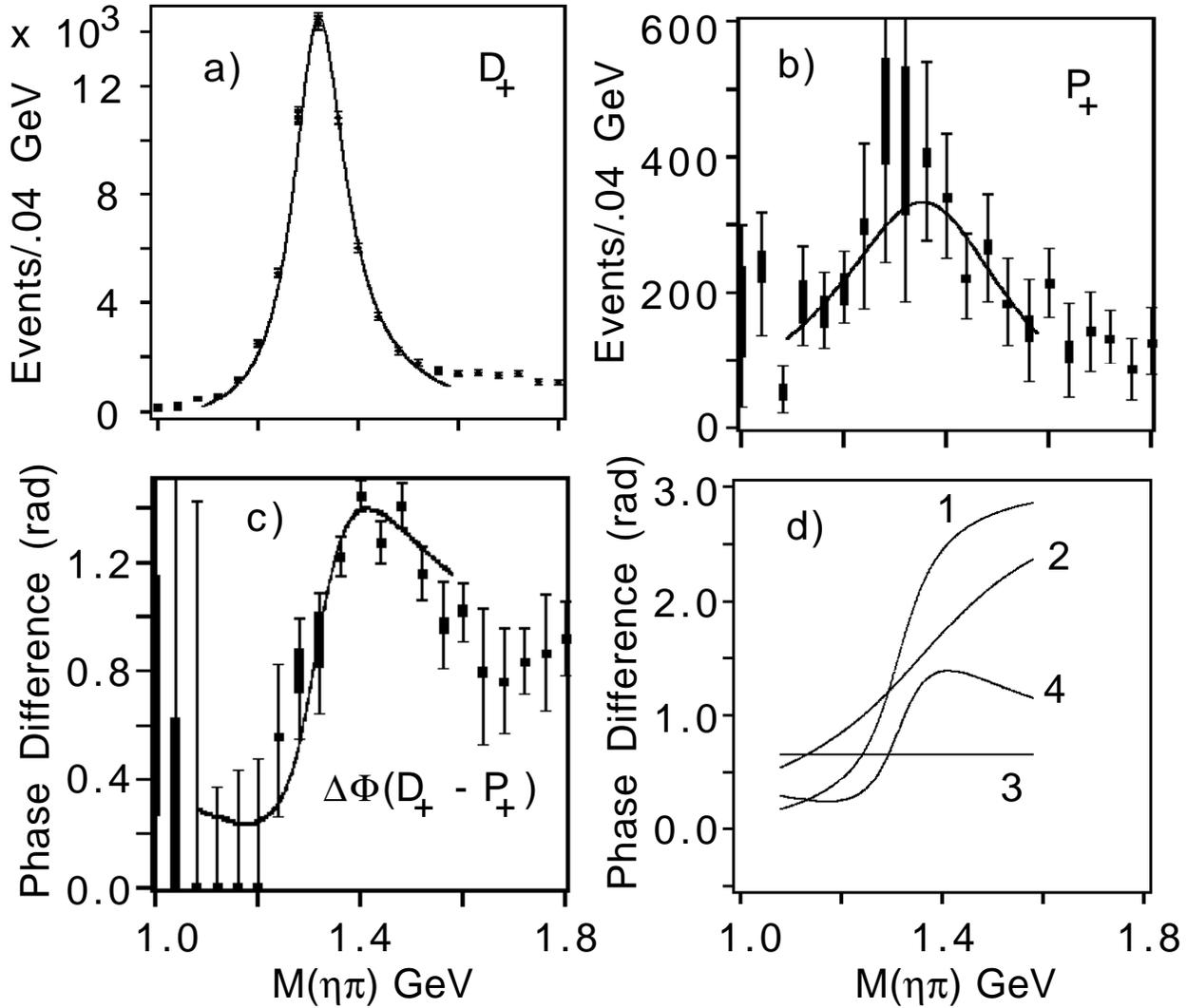,width=5.0in}}\vskip5mm
\end{center}
 \caption{Results of the partial wave amplitude analysis.  Shown are
a.) the fitted intensity distributions for the D$_{+}$ and b.)  the
P$_{+}$  partial waves, and c.) $\Delta\Phi({\rm D}_+-{\rm P}_+)$, their 
phase difference. The range of
values for the eight ambiguous solutions is shown by the central
bar and the  extent of the maximum error is shown by the error bars. 
Also shown as curves in a.), b.), and c.)
are the results of the mass dependent analysis described in the text.  
The lines in d.) correspond to (1) the fitted D$_{+}$ Breit-Wigner phase,
(2) the fitted P$_{+}$ Breit-Wigner phase, (3) the fitted D$_{+}-$P$_{+}$
relative production phase, 
and (4) the overall D$_{+}-$P$_{+}$ phase difference as shown in c.) but
with a different scale.}
\label{f:wave1}
\end{figure}

The results of the PWA fit in 40 ${\rm MeV}/c^{2}$ bins for $0.98 < 
M(\eta \pi^{-}) < 1.82$ ${\rm GeV}/c^{2}$ and
$0.10 < |t| < 0.95$ ${\rm GeV}^{2}$
are shown in Fig.\,\ref{f:wave1}a-c. Here, the
acceptance-corrected numbers 
of events predicted by the PWA fit for the D$_{+}$
and P$_{+}$ waves and their phase difference 
$\Delta\Phi({\rm D}_+-{\rm P}_+)$ are shown
as a function of $M(\eta \pi^{-})$.  
There are eight ambiguous solutions in the fit 
\cite{th:SUtwo,th:amb1,th:amb2}, each of which leads
to the same angular distribution.
We show the range of fitted values for these ambiguous
solutions in the vertical
rectangular bar at each mass bin, 
and the maximum extent of their errors is
shown as the error bar.
The $a_{2}(1320)$ is clearly
observed in the D$_{+}$ partial wave (Fig.\,\ref{f:wave1}a).  
A broad peak is seen 
in the  P$_{+}$ wave at about $1.4$ ${\rm GeV}/c^{2}$ (Fig.\,\ref{f:wave1}b).
$\Delta\Phi({\rm D}_+-{\rm P}_+)$ increases
through the $a_{2}(1320)$ region, and then
decreases above about 1.5 ${\rm GeV}/c^{2}$ (Fig.\,\ref{f:wave1}c).
The intensities for the  waves of 
negative reflectivity (not shown) are generally small and are all 
consistent with zero above about 1.3 ${\rm GeV}/c^{2}$.

These results are quite consistent with the VES
results\cite{ex:VESetapm}.  In particular, 
the shape of the phase difference
is virtually identical to that 
reported by  VES. (The magnitude of the phase difference is shifted
by about 20$^\circ$ relative to that of VES.)

Consistency checks and tests of the data have been carried 
out to determine whether the observation of the structure in the 
P$_{+}$ wave could be an artifact due to assumptions made in
the analysis or to
acceptance problems.  These include: fitting the data in restricted ranges
of the decay angle; inclusion of higher angular momentum states; fitting
the data with various t cuts; fitting the data using different parametrizations
of the background; making cuts on other kinematic variables such as the
$\pi^-p$ or the $\eta p$  effective masses; and fitting data using events
with $\eta\rightarrow\pi^+\pi^-\pi^0$ decays
(with rather different acceptance
from the 2$\gamma$ events).  
The results are
very stable and, in particular, the behavior of 
$\Delta\Phi({\rm D}_+-{\rm P}_+)$  does not
change  in any of these checks.  

Fits were also carried out on Monte Carlo events
generated with a pure D$_{+}$ wave to  determine whether 
P$_{+}$-wave structure could
be artificially induced by acceptance effects, resolution, or statistical
fluctuations.  We do find that some P$_{+}$ intensity can
be induced by resolution and/or acceptance effects. Such
``leakage" leads to a P$_{+}$ wave that
mimics the generated D$_{+}$ intensity (and in our case
would therefore have the shape of the $a_2(1320)$)
with a $\Delta\Phi({\rm D}_+-{\rm P}_+)$
that is independent of mass.  Neither property
is seen in our result.

In an attempt to understand the nature of the P$_{+}$  wave
observed in our experiment, we have carried out a 
\em mass-dependent \rm 
fit to the results of the mass-independent amplitude analysis.
The fit has been carried out in the $\eta\pi$ mass range
from 1.1 to 1.6 ${\rm GeV}/c^2$.  
The input quantities to the fit included, in each
mass bin, the
P$_{+}$-wave intensity; the D$_{+}$-wave intensity; and the 
D$_{+}-$P$_{+}$ phase difference.
Each of these quantities was taken with its error and correlation
coefficients
from the result of the amplitude analysis.
In this fit, we have assumed 
that the D$_{+}$-wave
and the P$_{+}$-wave decay amplitudes are  resonant and
have used relativistic Breit-Wigner forms\cite{BWnote} for 
these amplitudes. 
We introduce a constant relative production phase between the
P$_{+}$-wave and D$_{+}$-wave amplitudes.  The
parameters of the fit included the D$_{+}$-wave mass, width and intensity; 
the P$_{+}$-wave mass, width  and intensity; and
the  D$_{+}-$P$_{+}$ production phase difference.  
One can view this fit as a test of
the hypothesis that the correlation between the fitted P-wave intensity
and its phase (as a function of mass)
can be fit with a resonant Breit-Wigner amplitude.

Results of the 
fit are shown as the smooth curves in 
Fig.\,\ref{f:wave1}a, b, and c.  
The mass and width
of the $J^{PC}$ = $2^{++}$ state (Fig.\,\ref{f:wave1}a)  are 
(1317 $\pm1$ $\pm2$) ${\rm MeV}/c^2$ and 
(127 $\pm2$ $\pm2$) ${\rm MeV}/c^2$ respectively \cite{pdg}. 
(The first error given is statistical and the second
is systematic \cite{syst}.)
 The mass and width of the $J^{PC}$ = $1^{-+}$
state as shown in Fig.\,\ref{f:wave1}b 
are (1370  $\pm16$ ${+50}\atop{-30}$) ${\rm MeV}/c^2$ and 
(385  $\pm40$ ${+65}\atop{-105}$) ${\rm MeV}/c^2$ respectively.    Shown in
Fig.\,\ref{f:wave1}d are the Breit-Wigner phase dependences for the
$a_{2}(1320)$ (line 1) and the P$_+$ waves (line 2);
the fitted D$_{+}-$P$_{+}$ production phase difference (line 3);
and the fitted D$_{+}-$P$_{+}$ phase difference (line 4).  (Line 4,
which is identical to the fitted curve shown in Fig.\,\ref{f:wave1}c, is
obtained as line 1 $-$ line 2 $+$ line 3.)

The fit to the resonance hypothesis has a $\chi^2/$dof of 1.49.  The 
fact that the production phase difference can be fit by a 
mass-independent constant  (of 0.6 rad) is consistent with 
Regge-pole phenomenology \cite{regge2}
in the absence of final-state interactions.  
If one fits the data to a non-resonant (constant
phase) P$_+$ wave, and also assumes
a Gaussian intensity distribution
for the P$_+$ wave, one obtains a fit with
a $\chi^2/$dof of 1.55.  In this case,
the observed phase dependence on 
mass is attributed to
a rapidly varying production phase \cite{nonres}.  Such 
a phase variation cannot be excluded, but is not
expected for any known model.  Note that
for this non-resonant hypothesis 
one must  have a separate hypothesis
for the observed structure in
the P$_+$ intensity --- a structure which is 
explained naturally by the resonance hypothesis.
We thus conclude  that there is
credible evidence for the
production of a $J^{PC}$ = $1^{-+}$ exotic meson.

We would like to express our deep appreciation to the members of the MPS group.
Without their outstanding efforts, the results presented here could not have
been obtained.  
We  would also  like to acknowledge the 
invaluable assistance of the
staffs of the AGS and BNL, and of the various collaborating institutions.
This research was supported in part by the National Science Foundation,
the US Department of Energy, and the 
Russian State Committee for Science and
Technology.
 

%
\vfil\eject\vglue2cm
\vfil\eject\vglue5mm
 

\begin{references}
\bibitem{GJframe}The Gottfried-Jackson frame is a rest frame of the $\eta \pi^{-}$
system in which the z-axis is in the direction of 
the beam momentum, and the y-axis is in the direction of
the vector cross-product of the target and recoil momenta.
\bibitem{regge2} The
signature factor and the residue functions are at most 
t-dependent (not mass
dependent)  (see ref. \cite{th:regge}).
\end{references}
\end{document}